\documentclass[submission,copyright,creativecommons]{eptcs}
\providecommand{\event}{Non-Classical Logic. Theory and Applications} 
\usepackage{breakurl}             
\usepackage{underscore}           
\usepackage{amssymb}
\usepackage{mathtools}
\usepackage{xcolor}

\title{Mortensen Logics}
\author{Luis Estrada-González
	\institute{Institute for Philosophical Research\\
		National Autonomous University of Mexico\\
		Mexico City, Mexico}
	\institute{Faculty of Physico-Mathematical Sciences\\Meritorious Autonomous University of Puebla
		\thanks{This paper was written during the COVID-19 crisis. The authors want to thank the reviewers for their precious comments, as well as the support from the PAPIIT project IG400422 and from the Notre Dame International-Mexico Faculty Grant Program project ``The scope and limits of non-detachable conditionals''. The first author also wants to acknowledge the support from the DGAPA-UNAM through a PASPA sabbatical grant and from the Coimbra Group and the KU Leuven through a scholarship from the Programme for Young Professors and Researchers from Latin American Universities.}\\
		Puebla, Mexico}
	\email{loisayaxsegrob@comunidad.unam.mx}
	\and
	Fernando Cano-Jorge
	\institute{Universidad Panamericana\\
		Mexico City}
	\email{fernando.cano91@gmail.com}
}
\def\titlerunning{Mortensen Logics}
\def\authorrunning{L. Estrada-González \& F. Cano-Jorge}
\begin{document}
\maketitle

\begin{abstract}
In \cite{Mortensen1984}, Mortensen introduced a connexive logic commonly known as `\textbf{M3V}'. \textbf{M3V} is obtained by adding a special conditional to \textbf{LP}. Among its most notable features, besides its being connexive, \textbf{M3V} is negation-inconsistent and it validates the negation of every conditional. But Mortensen has also studied and applied extensively other non-connexive logics, for example, \emph{closed set logic}, \textbf{CSL}, and a variant of Sette's logic, identified and called `\textbf{P$^2$}' by Marcos in \cite{Marcos2006}.

In this paper, we analyze and compare systematically the connexive variants of \textbf{CSL} and \textbf{P$^2$}, obtained by adding the \textbf{M3V} conditional to them. Our main observations are two. First, that the inconsistency of \textbf{M3V} is exacerbated in the connexive variant of closed set logic, while it is attenuated in the connexive variant of the Sette-like \textbf{P$^2$}. Second, that the \textbf{M3V} conditional is, unlike other conditionals, \emph{connexively stable}, meaning that it remains connexive when combined with the main paraconsistent negations.
\end{abstract}

\section{Introduction}
In a paraconsistent context where formulas have three admissible assignments, and assuming the standard properties with respect to the ``classical'' assignments, that is

\begin{center}
    \begin{tabular}{c|c}
			$A$ & $N A$ \\
			\hline
			$\{1\}$ 	& $\{0\}$ \\
			$\{1, 0\}$ & \\
			$\{0\}$ 	& $\{1\}$ \\	
	\end{tabular}
\end{center}
		
\noindent
there are only two possibilities for paraconsistent negation $N$, namely the de Morgan negation found in González-Asenjo/Priest's \textbf{LP} and the negation of Sette's \textbf{P$^1$}, respectively:

\begin{center}
    \begin{tabular}{c|c}
			$A$ & $\sim \! A$ \\
			\hline
			$\{1\}$ 	& $\{0\}$ \\
			$\{1, 0\}$ & $\{1, 0\}$\\
			$\{0\}$ 	& $\{1\}$ \\	
	\end{tabular}
	\hfil
	\begin{tabular}{c|c}
			$A$ & $\neg A$ \\
			\hline
			$\{1\}$ 	& $\{0\}$ \\
			$\{1, 0\}$ & $\{1\}$\\
			$\{0\}$ 	& $\{1\}$ \\
	\end{tabular}
\end{center}

Among his many contributions in logic and philosophy, Chris Mortensen introduced a connexive logic commonly known as `\textbf{M3V}'. \textbf{M3V} is obtained by adding a special conditional to González-Asenjo/Priest's \textbf{LP}. Such conditional is structurally the same as the one used by Anderson and Belnap in \cite{AndersonBelnap1975} to show the consistency of the logic \textbf{E} and, in particular, to show how to block the paradox of necessity, i.e. to avoid validating formulas of the form $X>(Y>Z)$, with $>$ an entailment connective, $X$ a contingent truth and $(Y>Z)$ a logical truth.\footnote{A logic containing \textbf{M3V} was developed around the same time by Peña to cope with comparatives, gradables and vagueness. See \cite{Pena1995} for a summary of his results and \cite{Paoli2006} for a more friendly exposition of them.} Among its most notable features, besides its being connexive, \textbf{M3V} is negation-inconsistent and it validates the negation of every conditional.

But Mortensen has also studied and applied extensively other non-connexive logics. On the one hand there is \emph{closed set logic}, \textbf{CSL}, a paraconsistent logic motivated by dualizing open set logic, i.e. intuitionistic logic. \textbf{CSL} has notoriously been found defective in lacking a conditional connective because in it there is no connective $\copyright$ such that $A\copyright B$ is untrue if $A$ is true and $B$ untrue, as one would expect from a conditional. The two most obvious candidates, $\neg A\vee B$ and $\neg (A\wedge\neg B)$ are true when $A$ is true and $B$ is untrue, delivering thus countermodels to Detachment.\footnote{Mortensen has always argued that this is not a serious defect, especially when it comes to doing mathematics with \textbf{CSL}. We will not address this issue here. The fact is that there is no such connective in the logic; how bad is that is a different discussion.} On the other hand, in \cite{Mortensen1989a} he proposed another logic, which later Marcos \cite{Marcos2006} modified to obtain a variant of Sette's logic, identified and called \textbf{P$^2$} by Marcos.

In this paper, we analyze and compare systematically the connexive variants of \textbf{CSL} and \textbf{P$^2$}, obtained by adding the \textbf{M3V} conditional to them. Our main observations are two. First, that the inconsistency of \textbf{M3V} is exacerbated in the connexive variant of closed set logic, while it is attenuated in the connexive variant of the Sette-like \textbf{P$^2$}. Second, that the \textbf{M3V} conditional is, unlike other conditionals, \emph{connexively stable}, meaning that it validates the core connexive schemas when combined with the main paraconsistent negations.

The plan of the paper is as follows. In Section 2 we present some preliminary, general notions that will be useful for the remainder of the paper. In Section 3 we present \textbf{M3V} and mention some of its properties; some of them are already well-known, but others are noticed here for the first time. In Section 4 we introduce \textbf{cCSL3}, closed set logic restricted to three admissible interpretations, like \textbf{M3V}, enriched with the \textbf{E}-conditional. We give some of its most notable features, including likenesses and differences with \textbf{M3V}. There we show that, unlike other conditionals, the \textbf{E}-conditional is connexively stable with respect to both $\sim$ and $\neg$. Finally, in Section 5 we present \textbf{cP$^2$}. It shares the $\{\sim, \rightarrow_{\tiny{\textbf{E}}}\}$-fragment with \textbf{M3V}, but still they differ in ways that are significant for connexive logicians.

\section{Preliminary notions}
Let $A$ and $B$ arbitrary formulas of a given formal language, and $\Gamma$ a set of formulas of that language. In this paper, logical consequence is understood as truth-preservation from premises to conclusions in all interpretations, that is:

\begin{itemize}
\item[] $\Gamma\models_{\tiny{\textbf{L}}} A$ if and only if, for all $\sigma$, if $1\in\sigma(B)$ for all $B\in\Gamma$ then $1\in\sigma(A)$
\end{itemize}

\noindent
Now, let $N$ and $>$ be a negation and a conditional, respectively. \emph{Unrestricted Detachment} is logically valid in \textbf{L} iff

$$A, A> B\models_{\tiny{\textbf{L}}} B$$

\bigskip

\noindent
A logic \textbf{L} is \emph{connexive} iff the following hold:

\noindent
$\models_{\tiny{\textbf{L}}} N \! (A>N \! A)$\ \ \ \ \ \ \ \ \ \ \ \ \ \ \ \ \ \ \ \ \ \ \ Aristotle's Thesis

\noindent
$\models_{\tiny{\textbf{L}}} N \! (N \! A> A)$\ \ \ \ \ \ \ \ \ \ \ \ \ \ \ \ \ \ \ \ \ \ \ Variant of Aristotle's Thesis

\noindent
$\models_{\tiny{\textbf{L}}} (A > B) > N \! (A > N \! B)$\ \ \ \ \ \ Boethius' Thesis

\noindent
$\models_{\tiny{\textbf{L}}} (A > N \! B) > N \! (A > B)$\ \ \ \ \ \ Variant of Boethius' Thesis

\noindent
and

\noindent
$\not\models_{\tiny{\textbf{L}}}(A> B)>(B> A)$\ \ \ \ \ \ \ \ \ \ \ Non-symmetry of implication

\bigskip

\noindent
A logic \textbf{L} is \emph{hyper-connexive} iff it is connexive and at least one of the following holds:

\noindent
$\models_{\tiny{\textbf{L}}} N \! (A > N \! B) > (A > B)$\ \ \ \ \ \ Converse of Boethius' Thesis

\noindent
$\models_{\tiny{\textbf{L}}} N \! (A > B) > (A > N \! B)$\ \ \ \ \ \ Converse of Variant of Boethius' Thesis

\bigskip

\noindent
A logic \textbf{L} is \emph{nexive} iff the following hold:

\noindent
$\models_{\tiny{\textbf{L}}} N \! (A>N \! A)$\ \ \ \ \ \ \ \ \ \ \ \ \ \ \ \ \ \ \ \ \ \ \ Aristotle's Thesis

\noindent
$\models_{\tiny{\textbf{L}}} N \! (N \! A> A)$\ \ \ \ \ \ \ \ \ \ \ \ \ \ \ \ \ \ \ \ \ \ \ Variant of Aristotle's Thesis

\noindent
$\models_{\tiny{\textbf{L}}} (N \! A > B) > N \! (A > B)$\ \ \ \ \ \ Francez's Thesis

\noindent
$\models_{\tiny{\textbf{L}}} (A > B) > N \! (N \! A > B)$\ \ \ \ \ \ Variant of Francez's Thesis

\noindent
and

\noindent
$\not\models_{\tiny{\textbf{L}}}(A> B)>(B> A)$\ \ \ \ \ \ \ \ \ \ \ Non-symmetry of implication

\bigskip

\noindent
A logic \textbf{L} is \emph{hyper-nexive} iff it is nexive and at least one of the following holds:

\noindent
$\models_{\tiny{\textbf{L}}} N \! (A > B) > (N \! A > B)$\ \ \ \ \ \ Converse of Francez's Thesis

\noindent
$\models_{\tiny{\textbf{L}}} N \! (N \! A > B) > (A > B)$\ \ \ \ \ \ Converse of Boethius' Thesis

\bigskip

\noindent
A logic \textbf{L} is \emph{contradictory} or \emph{negation-inconsistent} iff there is an $A$ such that $\models_{\tiny{\textbf{L}}} A$ and $\models_{\tiny{\textbf{L}}} N \! A$.

\section{Mortensen's three-valued connexive logic}
The logic \textbf{M3V} was introduced, although not with that name, in \cite{Mortensen1984} (the name was given in \cite{McCall2012}, presumably to mean ``Mortensen's 3-valued connexive logic''). The following truth tables, with $V_{\tiny{\textbf{M3V}}}=\{2, 1, 0\}$ and $D^{+}=\{2, 1\}$, characterize \textbf{M3V}:

\begin{center}
	\begin{tabular}{cc|c|c|c|c}
		$A$ & $B$ & $\sim \! A$ & $A\wedge B$ & $A\vee B$ & $A\rightarrow_{\tiny{\textbf{E}}} B$\\
		\hline
		$2$	& $2$	&	$0$		& $2$		& $2$		& $1$\\
		$2$	& $1$	& 	$0$		& $1$		& $2$	    & $0$\\
		$2$	& $0$	& 	$0$		& $0$		& $2$		& $0$\\
		$1$	& $2$	& 	$1$		& $1$		& $2$		& $1$\\
		$1$ & $1$	& 	$1$		& $1$ 		& $1$		& $1$\\
		$1$ & $0$	& 	$1$		& $0$		& $1$		& $0$\\
		$0$	& $2$	& 	$2$		& $0$		& $2$		& $1$\\
		$0$	& $1$	& 	$2$		& $0$		& $1$		& $1$\\
		$0$	& $0$	& 	$2$		& $0$		& $0$		& $1$
	\end{tabular}
\end{center}

\noindent
A biconditional can be defined as usual, that is, as $(A\rightarrow_{\tiny{\textbf{E}}} B)\wedge(B\rightarrow_{\tiny{\textbf{E}}} A)$.

It must be noted that Mortensen's satisfiability conditions for the conditional are structurally the same as the ones used by Anderson and Belnap in \cite{AndersonBelnap1975} to show the consistency of the logic \textbf{E}, hence the subscript. In particular, they showed how to block the paradox of necessity, i.e. to avoid validating formulas of the form $X>(Y>Z)$, where $X$ is a contingent truth and $(Y>Z)$ is a logical truth.\footnote{A logic containing \textbf{M3V} was developed around the same time by Peña to cope with comparatives, gradables and vagueness. See \cite{Pena1995} for a summary of his results and \cite{Paoli2006} for a more friendly exposition of them.}

The three-valued nature of Mortensen's logic, along with the number of elements in $D^{+}$ and the evaluation conditions for negation motivate the representation of Mortensen's 2, 1, 0 as three subsets of the set of classical values $\{1, 0\}$, namely $\{1\}$, $\{1, 0\}$ and $\{0\}$, respectively, leaving the remaining subset $\{ \ \}$ aside as in the two-valued relational semantics for \textbf{LP}:

\begin{center}
\begin{tabular}{cc|c|c|c|c}
$A$ & $B$ & $\sim \! A$ & $A\wedge B$ & $A\vee B$ & $A\rightarrow_{\tiny{\textbf{E}}}B$\\
\hline
$\{1\}$ & $\{1\}$	& $\{0\}$	& $\{1\}$	& $\{1\}$	& $\{1, 0\}$\\
$\{1\}$ & $\{1, 0\}$	& $\{0\}$		& $\{1, 0\}$	& $\{1\}$		& $\{0\}$\\
$\{1\}$         & $\{0\}$		& $\{0\}$		& $\{0\}$		& $\{1\}$		& $\{0\}$\\
$\{1, 0\}$      & $\{1\}$		& $\{1, 0\}$	& $\{1, 0\}$	& $\{1\}$		& $\{1, 0\}$\\
$\{1, 0\}$      & $\{1, 0\}$	& $\{1, 0\}$	& $\{1, 0\}$  & $\{1, 0\}$    & $\{1, 0\}$\\
$\{1, 0\}$      & $\{0\}$		& $\{1, 0\}$	& $\{0\}$		& $\{1, 0\}$	& $\{0\}$\\
$\{0\}$          & $\{1\}$		& $\{1\}$		& $\{0\}$		& $\{1\}$		& $\{1, 0\}$\\
$\{0\}$          & $\{1, 0\}$	& $\{1\}$		& $\{0\}$		& $\{1, 0\}$	& $\{1, 0\}$\\
$\{0\}$          & $\{0\}$		& $\{1\}$		& $\{0\}$		& $\{0\}$		& $\{1, 0\}$
\end{tabular}
\end{center}

Applying the mechanical procedure described in \cite{OmoriandSano2015} for turning truth tables employing three of the four truth values of \textbf{FDE} into Dunn conditions (i.e., pairs of positive and negative conditions in terms of containing or not containing the classical values 0 or 1), we define a relation $\sigma$, which takes formulas as its domain and the set of truth values $\{1, 0\}$ as its codomain.

Then, the positive condition describes the cases in which $1\in\sigma(X)$, and the negative condition describes the cases in which $0\in\sigma(X)$. From the truth tables above we can infer that the conditions for the implication-free fragment of the language are standard, and that the clauses for $\rightarrow_{\tiny{\textbf{E}}}$ are as follows:    

\begin{itemize}
\item $1\in\sigma(A\rightarrow_{\tiny{\textbf{E}}} B)$ if a and only if $1\notin A$, or $0\notin B$, or both $0\in A$ and $1\in B$

\item $0\in\sigma(A\rightarrow_{\tiny{\textbf{E}}} B)$ if and only if $1\in\sigma(A)$ or $0\in\sigma(A)$ and either $1\in\sigma(B)$ or $0\in\sigma(B)$
\end{itemize}


\noindent
We are now in a position to point out some of \textbf{M3V}'s main features.

\begin{itemize}
\item Unlike \textbf{LP}, \textbf{M3V} validates unrestricted Detachment.
\item It is connexive.
\item It is contradictory. As witnesses, consider $(A\wedge\sim \! A)\rightarrow_{\tiny{\textbf{E}}} A$ and $\sim \! ((A\wedge\sim \! A)\rightarrow_{\tiny{\textbf{E}}} A)$.
\item All conditionals are false in \textbf{M3V}. The falsity condition for the conditional is but a sophisticated way of expressing $0\in\sigma(A\rightarrow_{\tiny{\textbf{E}}} B)$, which implies that $\models_{\tiny{\textbf{M3V}}} \sim \! (A \rightarrow_{\tiny{\textbf{E}}} B)$, for any $A$ and $B$.
\item Though all conditionals are false in \textbf{M3V}, some of them are true as well. Simply consider a conditional where both antecedent and consequent are just true. The conditional is false, yet true as well.
\item $\models_{\tiny{\textbf{M3V}}} \sim \! (A \rightarrow_{\tiny{\textbf{E}}} B)$ implies $\models_{\tiny{\textbf{M3V}}}\sim \! (A\rightarrow_{\tiny{\textbf{E}}} \sim \! B)$, by a simple substitution in the consequent. Due to the validity of the latter, we say that \textbf{M3V} is \emph{ultra-Abelardian}.\footnote{Claudio Pizzi has urged the connexive logic community not to multiply the principles with names of ancient philosophers. However, that plays a role in keeping a healthy logical memory. Peter Abelard held that conditionals express natures and that natures are characterized positively. For example, he believed that it would not be part of a human’s nature to not be a stone, although being an animal would be. (For details see \cite{Martin2004}.) Thus, for him, no conditional of the form $A\rightarrow\sim \! B$, where $A$ is necessarily positive ---that is, its main connective is not a negation--- and $\sim \! B$ is not a subformula of $A$, is true on pain of contradiction. Omitting the constraints on $A$ and $\sim \! B$ would lead to ultra-Abelardianism.}
\item Almost obvious given the validity of $\sim \! (A\rightarrow_{\tiny{\textbf{E}}} B)$, but even more overlooked, is the fact that \textbf{M3V} validates some schemas from Abelian logic, namely the \emph{Centering} principles\footnote{Nonetheless, it does not validate the \emph{Meyer-Slaney relativity axiom (schema)}, characteristic of purely implicative Abelian logics:

\noindent
$\not\models_{\textbf{\tiny{M3V}}}((A\rightarrow_{\tiny{\textbf{E}}} B)\rightarrow_{\tiny{\textbf{E}}} B)\rightarrow_{\tiny{\textbf{E}}} A$

\noindent
(For a countermodel, let $\sigma(A) = \{0\}$ and $\sigma(B) = \{1\}$.) The validity of $\sim \! (A\rightarrow_{\tiny{\textbf{E}}} A)$ demands moreover a comparison with Meyer and Martin's \textbf{SI$\sim$I} ---see \cite{MeyerMartin2019}---, where such schema is valid too. In that logic, $(C\rightarrow D)\rightarrow((A\rightarrow C)\rightarrow (A\rightarrow D))$ and $(A\rightarrow C)\rightarrow((C\rightarrow D)\rightarrow (A\rightarrow D))$, both object-language expressions of transitivity, are valid, but their negations are not. Nevertheless, since all conditionals are false in \textbf{M3V}, the negation of these forms of transitivity is valid as well.}:

\noindent
$\models_{\tiny{\textbf{M3V}}} \sim \! (A\rightarrow_{\tiny{\textbf{E}}} A)$

\noindent
$\models_{\tiny{\textbf{M3V}}} \sim \! (A\rightarrow_{\tiny{\textbf{E}}} A)\leftrightarrow_{\tiny{\textbf{E}}}(A\rightarrow_{\tiny{\textbf{E}}} A)$

(This provides other witnesses of negation-inconsistency, namely $A\rightarrow_{\tiny{\textbf{E}}} A$ and $\sim \! (A\rightarrow_{\tiny{\textbf{E}}} A)$. 

\item \textbf{M3V} is not hyper-connexive. Suppose it were, and that the Converse of Boethius hold. By ultra-Abelardianism and Detachment, $A\rightarrow B$ would be valid, but it is not. (A similar argument can be run using the Converse of the Variant of Boethius and the falsity of all conditionals.)
\item Francez's logics (see \cite{Francez2016}; see also \cite{Francez2019} and \cite{Francez2021}) have been the only recognized nexive logics so far. But \textbf{M3V} is nexive too, as a consequence of all negated conditionals being true. It is not hyper-nexive, though. (The proof is similar to the proof that it is not hyper-connexive.)
\end{itemize}

From the above, perhaps the most surprising feature is the fact that all conditionals are false in \textbf{M3V}. Indeed, one could argue that \textbf{M3V} is an interesting logic in so far as having arbitrary false conditionals, among many otherwise familiar properties, is an interesting feature for a logic to have. Nonetheless, this may require some philosophical elucidation.

The first thing to be said is that Cantwell's logic for conditional negation \textbf{CN} and \textbf{M3V} are inter-definable. In particular, the \textbf{E}-conditional is the contraposable conditional defined with the conditional in \textbf{CN}; see \cite{OmoriWansing2020}. Thus, one could attempt to build upon the intuitive features of \textbf{CN} to obtain some extra-logical support for \textbf{M3V}. True, the intuitiveness of the basic notions do not transfer immediately to the derived notions, but it could be a start. 

We do not follow that route, though. In our view, it is not unreasonable to have a logic in which all conditionals are false. On the one hand, tradition has it that certain syllogisms that are deemed valid often lack some tacit premise. For example, from ``Every human is mortal'' infer ``I am mortal'', where premise ``I am a human'' is tacit, i.e. it is a suppressed or unstated truth or piece of information not mentioned explicitly yet being part of the argument so that the conclusion indeed follows. This kind of argument is called \emph{enthymeme} by Aristotle (\emph{Rethoric}, 1357a16-21) and the implication relation between its premises and its conclusions is called \emph{enthymematic implication} by Sylvan \cite[p. 142]{Sylvan1989BG6}. Following this line of thought, \textbf{M3V} might be considered as a logic of enthymematic implication, i.e. as a logic about conditional arguments that strictly speaking are invalid, since they always lack some antecedent, premise or background information in order to hold (i.e. in order to entail the conclusion or consequent), but which may also be accepted as valid \emph{sotto voce}, \emph{prima facie} or \emph{ceteris paribus}.

On the other hand, connexive logic has been intimately attached to counterfactual notions from its very (contemporary) beginnings. (See \cite{Angell1962}.). This is relevant because, for example, Alan Hájek has long argued, in still unpublished but much read
work, for the idea that most counterfactuals are false. (See \cite{Hajek20XX}.) According to him, the indeterminism and indeterminacy associated with most counterfactuals entail their falsehood. Yet, counterfactual reasoning seems to play an important role in science, and ordinary speakers judge many counterfactuals that they utter to be true. Thus, \textbf{M3V} could be regarded as both a (zero-order) formalization of a radical version of Hájek's ideas on the falsity of counterfactual conditionals, while also capturing the idea that some of them need to be true.\footnote{There are of course many ways to address Hájek's challenge, and many of them that do not require a contradictory logic. Here we simply suggest that \textbf{M3V} can be taken as a formalization of a certain form of that debate. For another proposal in the connexive vicinity to address Hájek's challenge, see \cite{KapsnerOmori2017}.}

Finally, Meyer and Martin wanted to provide a logic for Aristotle's syllogistic, which was irreflexive. In their logic \textbf{SI$\sim$I}\footnote{They do not call it in that way, though. However, we simply indicate what further axiom schemas are added to the basis \textbf{S}, with `I' standing for $A\rightarrow A$, and `$\sim$I' for $\sim \! (A\rightarrow A)$.}, $A\rightarrow A$ was treated as a borderline case, both a fallacy with no valid instances (due to the irreflexivity of entailment) and a validity (because of the truth-preservation account of entailment), hence the validity of both $A\rightarrow A$ and $\sim \! (A\rightarrow A)$. One could explore the idea that implication or entailment are relations so demanding that no sentences can be ever in that relation, hence the validity of $\sim \! (A\rightarrow B)$. However, as in the case of \textbf{SI$\sim$I}, one could argue that, for theoretical simplicity, in this case, the functional approach, the truth of some instances of $A\rightarrow B$ are required as well.

We know that all what we have said is far from convincing. However, making a full case for the conceptual usefulness of \textbf{M3V} is beyond our aims. We merely expressed some ideas to take this logic as more than a mathematical curiosity.

\section{Connexive closed set logic}
The logic that we call `closed set logic' was introduced algebraically in \cite{Mortensen1995} and subsequently studied in \cite{Mortensen2000}, \cite{Mortensen2003} and \cite{Mortensen2007}.\footnote{Although the ideas underlying it are older, going back at least to \cite{McKinseyTarski1948}. The first systematic treatment of that logic on its own was the proof-theoretical analysis in \cite{Goodman1981}.} We focus here in the restriction to three interpretations, \textbf{CSL3}, defined by the following tables:

\begin{center}
\begin{tabular}{cc|c|c|c}
$A$ & $B$ & $\neg A$ & $A\wedge B$ & $A\vee B$\\
\hline
$\{1\}$ & $\{1\}$	& $\{0\}$	& $\{1\}$	& $\{1\}$\\
$\{1\}$ & $\{1, 0\}$	& $\{0\}$		& $\{1, 0\}$	& $\{1\}$\\
$\{1\}$ & $\{0\}$		& $\{0\}$		& $\{0\}$		& $\{1\}$\\
$\{1, 0\}$      & $\{1\}$		& $\{1\}$	& $\{1, 0\}$	& $\{1\}$\\
$\{1, 0\}$      & $\{1, 0\}$	& $\{1\}$	& $\{1, 0\}$  & $\{1, 0\}$\\
$\{1, 0\}$      & $\{0\}$		& $\{1\}$	& $\{0\}$		& $\{1, 0\}$\\
$\{0\}$          & $\{1\}$		& $\{1\}$		& $\{0\}$		& $\{1\}$\\
$\{0\}$          & $\{1, 0\}$	& $\{1\}$		& $\{0\}$		& $\{1, 0\}$\\
$\{0\}$          & $\{0\}$		& $\{1\}$		& $\{0\}$		& $\{0\}$
\end{tabular}
\end{center}

It is common wisdom that there is no conditional in \textbf{CSL3}. Consider a connective defined as follows:

$$A\rightarrow B\coloneqq \neg A\vee B \ (= \neg(A\wedge \neg B))$$

\noindent
This connective does not validate Detachment.\footnote{Although, in all fairness, it validates a restricted version, due to Beall \cite{Beall2011}, \cite{Beall2015} in the context of \textbf{LP}, namely, $A,~ A~\rightarrow~ B~\models_{\tiny{\textbf{CSL3}}}~ B\vee(A\wedge\neg A)$.} There are several ways to expand \textbf{CSL3} with a conditional connective that validates Detachment. In fact, 2$^4$ non-connexive conditionals could do the job; see \cite[p. 72]{CarnielliMarcos2002}.

But consider the expansion \textbf{cCSL3}, which adds the \textbf{E}-conditional, one of the 2$^4$ mentioned above, to \textbf{CSL3}. Let us point out some of \textbf{cCSL3}'s main features, starting with those involving just its $\{\neg, \rightarrow_{\tiny{\textbf{E}}}\}$-fragment.

\begin{itemize}
\item \textbf{cCSL3} validates unrestricted Detachment.
\item All conditionals are false in \textbf{cCSL3} and so $\models_{\tiny{\textbf{cCSL3}}} \neg(A\rightarrow_{\tiny{\textbf{E}}}B)$, for any $A$ and $B$.
\item It is connexive.
\item It follows that if \textbf{cCSL3} is connexive and all conditionals are false in it, \textbf{cCSL3} is contradictory, just as \textbf{M3V}. As witnesses, consider $A\rightarrow_{\tiny{\textbf{E}}}A$ and $\neg(A\rightarrow_{\tiny{\textbf{E}}}A)$.
\item \textbf{cCSL3} is ultra-Abelardian.
\item \textbf{cCSL3} does not validate exactly the same centering principles as \textbf{M3V}. One has

\noindent
$\models_{\tiny{\textbf{cCSL3}}} \neg(A\rightarrow_{\tiny{\textbf{E}}}A)$

\noindent
$\models_{\tiny{\textbf{cCSL3}}} (A\rightarrow_{\tiny{\textbf{E}}}A)\rightarrow_{\tiny{\textbf{E}}}\neg(A\rightarrow_{\tiny{\textbf{E}}}A)$

\noindent
but also

\noindent
$\not\models_{\tiny{\textbf{cCSL3}}} \neg(A\rightarrow_{\tiny{\textbf{E}}}A)\rightarrow_{\tiny{\textbf{E}}}(A\rightarrow_{\tiny{\textbf{E}}}A)$

\item The above implies that \textbf{cCSL3} also lacks the Deduction Property. In fact, every logical truth in \textbf{cCSL3} entails any other logical truth, in particular, $\neg(A\rightarrow_{\tiny{\textbf{E}}}A)\models_{\tiny{\textbf{cCSL3}}}A\rightarrow_{\tiny{\textbf{E}}}A$, yet $\not\models_{\tiny{\textbf{cCSL3}}} \neg(A\rightarrow_{\tiny{\textbf{E}}}A)\rightarrow_{\tiny{\textbf{E}}}(A\rightarrow_{\tiny{\textbf{E}}}A)$.

\item The invalidity of $\neg(A\rightarrow_{\tiny{\textbf{E}}}A)\rightarrow_{\tiny{\textbf{E}}}(A\rightarrow_{\tiny{\textbf{E}}}A)$ generalizes. Since any conditional of the form $X\rightarrow_{\tiny{\textbf{E}}}Y$ is false and any conditional of the form $\neg(W\rightarrow_{\tiny{\textbf{E}}}Z)$ is just true, it follows that no conditional of the form $\neg(W\rightarrow_{\tiny{\textbf{E}}}Z)\rightarrow_{\tiny{\textbf{E}}}(X\rightarrow_{\tiny{\textbf{E}}}Y)$ is valid.\footnote{And the validity of $(A\rightarrow_{\tiny{\textbf{E}}}A)\rightarrow_{\tiny{\textbf{E}}}\neg(A\rightarrow_{\tiny{\textbf{E}}}A)$ also generalizes: every conditional of the form $(X\rightarrow_{\tiny{\textbf{E}}}Y)\rightarrow_{\tiny{\textbf{E}}}\neg(W\rightarrow_{\tiny{\textbf{E}}}Z)$ is valid.}

\item It is clear now that \textbf{cCSL3} and \textbf{M3V} validate different arguments. As another witness, consider $A\models_{\textbf{M3V}} \sim\sim \! A$ but $A\not\models_{\textbf{cCSL3}}\neg\neg A$.

\item \textbf{cCSL3} is not hyper-connexive. The argument is as for \textbf{M3V}.

\item \textbf{cCSL3} is nexive, just as \textbf{M3V}. And like \textbf{M3V}, it is not hyper-nexive. Again, the proof is an adaptation of the proof that \textbf{M3V} is not hyper-connexive.
\end{itemize}

A natural question at this point is whether $\neg$ is definable in \textbf{M3V}. It is not. It could be defined as $\sim \! \circ(A\rightarrow_{\tiny{\textbf{E}}}\sim \! \circ\circ \! A)$, with $\circ$ a consistency connective:

\begin{center}
    \begin{tabular}{c|c}
			$A$ & $\circ A$ \\
			\hline
			$\{1\}$ 	& $\{1\}$ \\
			$\{1, 0\}$ & $\{0\}$\\
			$\{0\}$ 	& $\{1\}$ \\	
	\end{tabular}
\end{center}

\noindent
But such a connective is not definable in \textbf{M3V}: The connective is not definable in \textbf{CN} as per \cite{Omori2016}, and a connective is definable in \textbf{M3V} iff it is definable in \textbf{CN}, as per \cite{OmoriWansing2020}.\footnote{What about defining the \textbf{LP} negation in \textbf{cCSL3}? We do not know, but our guess is that it cannot be defined.}

The list of properties above does not highlight enough some features of \textbf{cCSL3}, especially around connexive principles:

\begin{itemize}
\item $\neg(A\rightarrow_{\tiny{\textbf{E}}}B)$ is just true in all interpretations in \textbf{cCSL3}; $\sim \! (A\rightarrow_{\tiny{\textbf{E}}}B)$ is true in all interpretations in \textbf{M3V}, but it is also false under some of them. This has consequences for the connexive principles, as we will see.
\item Recall that, in \textbf{M3V}, Aristotle's Theses are true under all interpretations, although there are some interpretations under which they are also false. That is not the case in \textbf{cCSL3}: Aristotle's Theses are just true.
\item Boethius' Theses are true under all interpretations in \textbf{M3V}, although they are also false under all interpretations. That is the case as well in \textbf{cCSL3}, with the difference that in this logic, the negations of Boethius' Theses are just true.
\item Both $(A\wedge\sim \! A)\rightarrow_{\tiny{\textbf{E}}} A$ and $\sim \! ((A\wedge\sim \! A)\rightarrow_{\tiny{\textbf{E}}} A)$ are valid in \textbf{M3V}, they are both true and false in all interpretations. But although both $(A\wedge\neg A)\rightarrow_{\tiny{\textbf{E}}} A$ and $\neg((A\wedge\neg A)\rightarrow_{\tiny{\textbf{E}}} A)$ are valid in \textbf{cCSL3}, the latter is just true in all interpretations.
\item More generally: If both $X$ and $\sim \! X$ are valid in \textbf{M3V}, then $\neg X$ is just true in \textbf{cCSL3}, unlike $\sim \! X$ in \textbf{M3V}, even if $X$ fails to be valid in \textbf{cCSL3}. (The proof is straightforward. For schemas exemplifying this, recall the ones for the failure of the Deduction Property.)
\end{itemize}

Finally, an attractive feature of the \textbf{E}-conditional should be mentioned: unlike some well-known connexive conditionals in the literature, it is stable under changes of negation. Let us make that more precise.

Let a \emph{standard negation} be a unary connective $N$ satisfying that $\sigma(N A) = \{1\}$ if $\sigma(A) = \{0\}$, and $\sigma(N A) = \{0\}$ if $\sigma(A) = \{1\}$. If a standard negation $N$ is such that, in a logic \textbf{L}, $A, N A\not\models_{\tiny{\textbf{L}}} B$, we will call it a \emph{standard paraconsistent negation}. Let us define the \emph{type of standard paraconsistent negations} (TSPN) as the set of all such connectives definable according to a set of admissible evaluations. In the present context, TSPN only has two connectives: $\sim$ and $\neg$.

Now, let us say that a conditional $A>B$ is \emph{connexively stable} with respect to TSPN iff

\noindent
$\models_{\tiny{\textbf{L}}} N_i \! (A> N_i \! A)$

\noindent
$\models_{\tiny{\textbf{L}}} N_i \! (N_i \! A>A)$

\noindent
$\models_{\tiny{\textbf{L}}} (A>B)>N_i \! (A> N_i \! B)$

\noindent
$\models_{\tiny{\textbf{L}}} (A> N_i \! B)> N_i \! (A>B)$

\noindent
and

\noindent
$\not\models_{\tiny{\textbf{L}}}(A>B)>(B>A)$

\noindent
for each $N_i$ in TSPN. From the previous discussion, $\rightarrow_{\tiny{\textbf{E}}}$ is connexively stable with respect to TSPN. However, the main connexive conditionals in the literature are not connexively stable. The conditionals defined by the following tables validate the connexive principles only with $\sim$, but not with $\neg$:

\begin{center}
    \begin{tabular}{c|ccc}
        $A\rightarrow_{W} B$ & $\{1\}$ & $\{1, 0\}$ & $\{0\}$\\
        \hline
        $\{1\}$ 	& $\{1\}$       & $\{1, 0\}$   	& $\{0\}$\\
        $\{1, 0\}$  & $\{1\}$     	& $\{1, 0\}$ 	& $\{0\}$\\
        $\{0\}$ 	& $\{1, 0\}$ 	& $\{1, 0\}$ 	& $\{1, 0\}$\\
    \end{tabular}
\end{center}

\begin{center}
    \begin{tabular}{c|ccc}
        $A\rightarrow_{BL} B$ & $\{1\}$ & $\{1, 0\}$ & $\{0\}$\\
        \hline
        $\{1\}$ 	& $\{1\}$       & $\{0\}$    	& $\{0\}$\\
        $\{1, 0\}$  & $\{1\}$     	& $\{1, 0\}$ 	& $\{0\}$\\
        $\{0\}$ 	& $\{1, 0\}$ 	& $\{1, 0\}$ 	& $\{1, 0\}$\\
    \end{tabular}
\end{center}

\noindent
They are, respectively, Wansing's conditional from \cite{Wansing2005} restricted to three admissible interpretations ---found explicitly for three interpretations in \cite{Olkhovikov2001}, \cite{Cantwell2008}, \cite{Omori2016}---, and Belikov and Loginov's conditional from \cite{BelikovandLoginov20XX}. Although Aristotle's Thesis becomes just true under all interpretations with the first conditional and $\neg$, Boethius' Thesis fails: it is just false when $A$ is just true and $B$ is both true and false. The problem with the second conditional is a sort of dual: Boethius' Thesis is valid, but Aristotle's Thesis fails when $A$ is both true and false.

Note that Francez's conditional, from \cite{Francez2019}, restricted to three admissible interpretations, i.e.

\begin{center}
    \begin{tabular}{c|ccc}
        $A\rightarrow_{F} B$ & $\{1\}$ & $\{1, 0\}$ & $\{0\}$\\
        \hline
        $\{1\}$ 	& $\{1, 0\}$    & $\{1, 0\}$    & $\{0\}$\\
        $\{1, 0\}$  & $\{1, 0\}$ 	& $\{1, 0\}$ 	& $\{0\}$\\
        $\{0\}$ 	& $\{1, 0\}$ 	& $\{1, 0\}$ 	& $\{1, 0\}$\\
    \end{tabular}
\end{center}

\noindent
is also stable with respect to standard paraconsistent negations. It can be easily verified that it does not allow for the countermodels present in the previous conditionals.\footnote{Angell-McCall's conditional, found in \cite{Angell1962} and \cite{McCall1966}, is connexive with respect to Boolean negation, but not with respect to other standard explosive negations. The definition of this notion, and the verification of the claim about the Angell-McCall's conditional are left as an exercise.}

\section{Connexive P$^2$}
There is one more logic due partly to Mortensen, but also partly to Marcos. In \cite{Mortensen1989a}, Mortensen introduced a logic called `\textbf{C$_{0.2}$}' characterized by the following tables:

\begin{center}
\begin{tabular}{cc|c|c|c|c}
$A$ & $B$ & $\sim \! A$ & $A\wedge_{\tiny{\textbf{P}}} B$ & $A\vee_{\tiny{\textbf{P}}} B$ & $A\rightarrow_{\tiny{\textbf{P}}}B$\\
\hline
$\{1\}$     & $\{1\}$	    & $\{0\}$	    & $\{1\}$	& $\{1\}$	& $\{1\}$\\
$\{1\}$     & $\{ \ \}$	& $\{0\}$		& $\{1\}$	& $\{1\}$	& $\{1\}$\\
$\{1\}$     & $\{0\}$		& $\{0\}$		& $\{0\}$	& $\{1\}$	& $\{0\}$\\
$\{ \ \}$  & $\{1\}$		& $\{ \ \}$	    & $\{1\}$	& $\{1\}$	& $\{1\}$\\
$\{ \ \}$  & $\{ \ \}$	& $\{ \ \}$	    & $\{1\}$   & $\{1\}$   & $\{1\}$\\
$\{ \ \}$  & $\{0\}$		& $\{ \ \}$	    & $\{0\}$	& $\{1\}$	& $\{0\}$\\
$\{0\}$     & $\{1\}$		& $\{1\}$		& $\{0\}$	& $\{1\}$	& $\{1\}$\\
$\{0\}$     & $\{ \ \}$	& $\{1\}$		& $\{0\}$	& $\{1\}$	& $\{1\}$\\
$\{0\}$     & $\{0\}$		& $\{1\}$		& $\{0\}$	& $\{0\}$	& $\{1\}$
\end{tabular}
\end{center}

\noindent
(Mortensen originally used three values, 1, 2, and 3, being 1 the only designated value. We are taking advantage here of the Dunn semantics, as mentioned in Section 3.)

Marcos \cite{Marcos2006} suggested to replace the interpretation $\{ \ \}$ by the interpretation $\{1, 0\}$ ---or, in his original terms, to make the value 2 designated along with 1---, and put $\sim$ instead of $\neg$, to get the logic \textbf{P$^2$}, whose tables look like these:

\begin{center}
\begin{tabular}{cc|c|c|c|c}
$A$ & $B$ & $\sim \! A$ & $A\wedge_{\tiny{\textbf{P}}} B$ & $A\vee_{\tiny{\textbf{P}}} B$ & $A\rightarrow_{\tiny{\textbf{P}}}B$\\
\hline
$\{1\}$     & $\{1\}$	    & $\{0\}$	    & $\{1\}$	& $\{1\}$	& $\{1\}$\\
$\{1\}$     & $\{1, 0\}$	& $\{0\}$		& $\{1\}$	& $\{1\}$	& $\{1\}$\\
$\{1\}$     & $\{0\}$		& $\{0\}$		& $\{0\}$	& $\{1\}$	& $\{0\}$\\
$\{1, 0\}$  & $\{1\}$		& $\{1, 0\}$	    & $\{1\}$	& $\{1\}$	& $\{1\}$\\
$\{1, 0\}$  & $\{1, 0\}$	& $\{1, 0\}$	    & $\{1\}$   & $\{1\}$   & $\{1\}$\\
$\{1, 0\}$  & $\{0\}$		& $\{1, 0\}$	    & $\{0\}$	& $\{1\}$	& $\{0\}$\\
$\{0\}$     & $\{1\}$		& $\{1\}$		& $\{0\}$	& $\{1\}$	& $\{1\}$\\
$\{0\}$     & $\{1, 0\}$	& $\{1\}$		& $\{0\}$	& $\{1\}$	& $\{1\}$\\
$\{0\}$     & $\{0\}$		& $\{1\}$		& $\{0\}$	& $\{0\}$	& $\{1\}$
\end{tabular}
\end{center}

Now, to get a connexive variant of this, \textbf{cP$^2$}, replace the \textbf{P}-conditional with the \textbf{E}-conditional, i.e. obtain a logic characterized by the following tables:

\begin{center}
\begin{tabular}{cc|c|c|c|c}
$A$ & $B$ & $\sim \! A$ & $A\wedge_{\tiny{\textbf{P}}} B$ & $A\vee_{\tiny{\textbf{P}}} B$ & $A\rightarrow_{\tiny{\textbf{E}}}B$\\
\hline
$\{1\}$     & $\{1\}$	    & $\{0\}$	    & $\{1\}$	& $\{1\}$	& $\{1, 0\}$\\
$\{1\}$     & $\{1, 0\}$	& $\{0\}$		& $\{1\}$	& $\{1\}$	& $\{0\}$\\
$\{1\}$     & $\{0\}$		& $\{0\}$		& $\{0\}$	& $\{1\}$	& $\{0\}$\\
$\{1, 0\}$  & $\{1\}$		& $\{1, 0\}$	& $\{1\}$	& $\{1\}$	& $\{1, 0\}$\\
$\{1, 0\}$  & $\{1, 0\}$	& $\{1, 0\}$	& $\{1\}$   & $\{1\}$   & $\{1, 0\}$\\
$\{1, 0\}$  & $\{0\}$		& $\{1, 0\}$	& $\{0\}$	& $\{1\}$	& $\{0\}$\\
$\{0\}$     & $\{1\}$		& $\{1\}$		& $\{0\}$	& $\{1\}$	& $\{1, 0\}$\\
$\{0\}$     & $\{1, 0\}$	& $\{1\}$		& $\{0\}$	& $\{1\}$	& $\{1, 0\}$\\
$\{0\}$     & $\{0\}$		& $\{1\}$		& $\{0\}$	& $\{0\}$	& $\{1, 0\}$
\end{tabular}
\end{center}

\textbf{M3V} and \textbf{cP$^2$} coincide in the $\{\sim, \rightarrow_{\tiny{\textbf{E}}}\}$-fragment, but they differ in ways that are significant for connexive logicians. Consider the following (non-core) connexive principles:

\noindent
$\sim \! ((A \rightarrow_{\tiny{\textbf{E}}} B) \wedge_{\tiny{\textbf{P}}} (\sim \! A \rightarrow_{\tiny{\textbf{E}}} B))$ \ \ \ \ \ \ Aristotle's Second Thesis

\noindent
$\sim \! ((A \rightarrow_{\tiny{\textbf{E}}} B) \wedge_{\tiny{\textbf{P}}} (A \rightarrow_{\tiny{\textbf{E}}} \sim \! B))$ \ \ \ \ \ \ \ Abelard's Principle

\noindent
These are valid in \textbf{M3V}, as originally reported in \cite{EstradaRamirez2016}, but they are not in \textbf{cP$^2$}. For a countermodel, consider the case when both $A$ and $B$ are both true and false. (This will do for both principles.) For the record, these are countermodels for the principles written in the language of \textbf{cCSL3}.

\paragraph{Short digression.} \noindent
There is a different, more direct way of presenting \textbf{cP$^2$}, starting directly with Sette's \textbf{P$^1$} without the detour through Mortensen's \textbf{C$_{2.0}$}. Consider Sette's logic \textbf{P$^1$}, characterized by the following truth tables:

\begin{center}
\begin{tabular}{cc|c|c|c|c}
$A$ & $B$ & $\neg A$ & $A\wedge_{\tiny{\textbf{P}}} B$ & $A\vee_{\tiny{\textbf{P}}} B$ & $A\rightarrow_{\tiny{\textbf{P}}}B$\\
\hline
$\{1\}$     & $\{1\}$	    & $\{0\}$	    & $\{1\}$	& $\{1\}$	& $\{1\}$\\
$\{1\}$     & $\{1, 0\}$	& $\{0\}$		& $\{1\}$	& $\{1\}$	& $\{1\}$\\
$\{1\}$     & $\{0\}$		& $\{0\}$		& $\{0\}$	& $\{1\}$	& $\{0\}$\\
$\{1, 0\}$  & $\{1\}$		& $\{1\}$	    & $\{1\}$	& $\{1\}$	& $\{1\}$\\
$\{1, 0\}$  & $\{1, 0\}$	& $\{1\}$	    & $\{1\}$   & $\{1\}$   & $\{1\}$\\
$\{1, 0\}$  & $\{0\}$		& $\{1\}$	    & $\{0\}$	& $\{1\}$	& $\{0\}$\\
$\{0\}$     & $\{1\}$		& $\{1\}$		& $\{0\}$	& $\{1\}$	& $\{1\}$\\
$\{0\}$     & $\{1, 0\}$	& $\{1\}$		& $\{0\}$	& $\{1\}$	& $\{1\}$\\
$\{0\}$     & $\{0\}$		& $\{1\}$		& $\{0\}$	& $\{0\}$	& $\{1\}$
\end{tabular}
\end{center}

To obtain \textbf{P$^2$}, simply replace $\neg$ by $\sim$, as it has been already noticed in \cite{Karpenko1999}.\footnote{He anticipated thus Marcos' formulation of \textbf{P$^2$}. However, Karpenko  wrongly claims that Mortensen's original logic \textbf{C$_{0.2}$} is paraconsistent. Karpenko assumed that the value 2, in Mortensen's presentation, is designated, which is not. Marcos correctly noticed that \textbf{P$^2$} requires a certain amount of dualization in Mortensen's \textbf{C$_{0.2}$}.} Then, to get \textbf{cP$^2$}, replace the \textbf{P}-conditional with the \textbf{E}-conditional.

\section{Conclusions}
In this paper, we took two interests of Mortensen, connexivity and certain brands of paraconsistency, and combined them into single logics. Although connexivity is at least a matter of two, negation and the conditional, the \textbf{E}-conditional of Mortensen's \textbf{M3V} excels among other conditional in validating the connexive schemas even when combined with other (paraconsistent) negations. Also, some features of \textbf{M3V} are exacerbated when a different negation is used. For example, in \textbf{M3V} all negated conditionals are true, but also sometimes false, whereas changing the negation leads to the result that negated conditionals are just true, never false.

There are at least five ways in which this work can be continued. First, when presented in slightly different ways, a logic might exhibit more interesting features. As we mentioned, the logics \textbf{CN} and \textbf{M3V} are inter-definable; it would be worth take a look at the logics defined here with other conditionals definable on them. Second, one could enrich the languages here with consistency connectives to make a comparison with the \textbf{LFIs}. Third, one could try to get both negations in a single language and study the effect of that on connexive principles. Fourth, at least in the \textbf{E}-conditional , Mortensen suggested to couple closed set logic with different notions of logical consequence. This would allow, among other things, to discriminate between schemas that are true under all interpretations and those that are just true under all interpretations. This would give rise to ``exactly true'' or ``non-falsity'' versions of the logics above, which have been studied in the vicinity of \textbf{FDE}. (See for example \cite{PietzRivieccio2013} and \cite{ShramkoZaitsevBelikov2019}, but also \cite{EstradaGonzalez2020} for a discussion closer to the present context.) Speaking of that, and finally, one can move the entire discussion on top of \textbf{FDE} to work with more admissible interpretations. That would augment the number of logical and conceptual distinctions available to work with.

\bibliographystyle{eptcs}
\bibliography{biblio}
\end{document}